\documentclass[12pt,preprint]{aastex}
\usepackage{epsfig,lscape}
\def\lap{\lower.5ex\hbox{$\; \buildrel < \over \sim \;$}}
\def\gap{\lower.5ex\hbox{$\; \buildrel > \over \sim \;$}}

\def\ergcm2s{${\rm eg\ cm^{-2}\ s^{-1}}$}

\def\ergscm2s{${\rm erg\ cm^{-2}\  s^{-1}}$}

\def\cm-2{${\rm cm^{-2}}$}

\begin{document}

\title{The Chandra ACIS Survey of M33 (ChASeM33):  Transient X-ray Sources Discovered in M33}

\author{Benjamin F. Williams\altaffilmark{1},
Terrance J. Gaetz \altaffilmark{2},
Frank Haberl,
Wolfgang Pietsch\altaffilmark{3},
Avi Shporer\altaffilmark{4},
Parviz Ghavamian\altaffilmark{5},
Paul P. Plucinsky\altaffilmark{2},
T. Mazeh\altaffilmark{4}, 
Manami Sasaki\altaffilmark{2},
and Thomas G. Pannuti\altaffilmark{6}
}
\altaffiltext{1}{University of Washington Astronomy Department, Box 351580, Seattle, WA, 98195; ben@astro.washington.edu}

\altaffiltext{2}{Harvard-Smithsonian Center for Astrophysics, 60 Garden Street, Cambridge, MA 02138; gaetz@ead.cfa.harvard.edu; plucinsk@head.cfa.harvard.edu; msasaki@cfa.harvard.edu}

\altaffiltext{3}{Max-Planck-Institut f\"ur extraterrestrische Physik, 85741 Garching, Germany; fwh@mpe.mpg.de; wnp@mpe.mpg.de}

\altaffiltext{4}{Wise Observatory, Raymond and Beverly Sackler Faculty of Exact
Sciences, Tel Aviv University, Tel Aviv 69978, Israel; shporer@wise.tau.ac.il; mazeh@wise.tau.ac.il}

\altaffiltext{5}{Department of Physics and Astronomy, Johns Hopkins University, 3400 North Charles Street, Baltimore, MD 21218; parviz@pha.jhu.edu}

\altaffiltext{6}{Space Science Center, Morehead State University, 200A Chandler Place, Morehead, KY 40351; t.pannuti@morehead-st.edu}

\keywords{X-rays: binaries --- galaxies: individual (M33) ---
binaries: close --- X-rays: stars}

\begin{abstract}

The {\it Chandra} ACIS Survey of M33 (ChASeM33) has acquired 7 fields
of ACIS data covering M33 with 200 ks of exposure in each field.  A
catalog from the first 10 months of data, along with archival {\it
Chandra} observations dating back to the year 2000, is currently
available. We have searched these data for transient sources that are
measured to have a 0.35-8.0 keV unabsorbed luminosity of at least
4$\times$10$^{35}$ erg s$^{-1}$ in one epoch and are not detected in
another epoch.  This set of the survey data has yielded seven such
sources, including one previously-known supersoft source.  We analyzed
{\sl XMM-Newton} data from the archive distributed over the years 2000
to 2003 to search for recurrent outbursts and to get a spectrum for
the supersoft transient. We find only one recurrent transient in our
sample.  The X-ray spectra, light curves, and optical counterpart
candidates of two of the other sources suggest that they are high-mass
X-ray binaries.  Archival {\sl Spitzer} photometry and high X-ray
absorption suggest that one of the sources is a highly variable
background active galactic nucleus.  The other three sources are more
difficult to classify.  The bright transient population of M33 appears
to contain a large fraction of high-mass X-ray binaries compared with
the transient populations of M31 and the Galaxy, reflecting the later
morphology of M33.

\end{abstract}

\section{Introduction}

X-ray transient sources are rare and valuable laboratories for the
study of accretion physics.  Most low-mass X-ray binaries that are
observed as transient X-ray sources have been shown to contain black
hole primaries \citep{mcclintock2006}.  These objects, known as black
hole binaries (BHBs), are of great interest for studies of disk
accreton in the strong gravity regime.  Other transient events come
from supersoft sources (SSSs; \citealp[e.g.][and references
therein]{distefano2004}), which are likely to harbor white dwarf
primaries \citep{king2002}.  Still other transient events occur in
high-mass X-ray binaries (HMXBs); these have hard spectra and often
have neutron star primaries. Most such systems are $Be$ transients
driven by wind-fed accretion \citep{tanaka1996}.

While there has been a significant amount of research dedicated to
observing and understanding the transient X-ray sources discovered in
the Galaxy \citep[][and many
others]{mcclintock2006,jain2001l,tomsick2005}, relatively little work
has been done to compare this population of exotic sources to their
extragalactic analogs.  Most of the existing extragalactic studies
concentrate on transients in the Magellanic Clouds
\citep{coe2001,kahabka1996} and M31 \citep[see][and references
therein]{williams2006}.  Studies of M31 suggest that the black
hole/neutron star (BH/NS) ratio in M31 could be higher than expected
from simple stellar evolution and that the transient rate in M31 is
comparable to or slightly higher than that of the Galaxy
\citep{williams2004}.  The Magellanic Clouds do not contain large
samples of LMXB transient sources, but the Small Magellanic Cloud
contains a large number of $Be$ binaries for its size
\citep{haberl2004}.

Because of its excellent spatial resolution ($\sim$1$''$ FWHM at 1
keV) and its ability to revisit the same part of the sky at several
different times of the year, the {\sl Chandra} X-ray Observatory is
particularly well-suited to conduct searches for transient X-ray
sources in nearby galaxies. Based on its low distance modulus of
$(m-M)_0=24.50$ \citep{vandenberg1991} and low inclination angle of 56
degrees \citep{zaritsky1989}, the nearby spiral galaxy M33 is a prime
target for such searches.  Compared to more distant spiral galaxies or
larger nearby spirals, M33 is relatively easy to fully monitor with
{\it Chandra}, allowing classification of the transient sources as
well as comparisons of the M33 transient population with those of the
Galaxy and M31.

Our {\it Chandra} Very Large Program (VLP) to survey M33 to a depth of
$\sim$5$\times$10$^{34}$ erg s$^{-1}$ with the ACIS-I array
(\citealp{pietsch2006,gaetz2007,plucinsky2008}; , ApJS, submitted) is
providing a wealth of data covering nearly all of the M33 disk at a
variety of epochs.  While we were not able to control the depth and
coverage of these epochs to optimize the data set for creating an
unbiased sample of X-ray transients, our data are still extremely
useful for discovering new transient sources.  We therefore have
carefully searched our first source catalog \citep{plucinsky2008} and
the \citet{grimm2005} {\sl Chandra} catalog for the brightest
transient sources in order to provide some detailed information about
the characteristics of transient X-ray sources in M33.

In this paper, we announce the discovery that six of the new X-ray
sources cataloged by the ChASeM33 survey are transient sources, and we
discuss our analysis of a seventh, previously-known transient.  Two of
the candidates have optical counterpart candidates with the color and
brightness of upper main sequence stars in M33, making them likely to
be high-mass X-ray binaries (HMXBs).  One candidate has a very red
counterpart, suggesting that it is a low-mass X-ray binary. Two more
have faint optical counterpart candidates, and the rest have no
detected optical counterparts.

\section{Data Analysis}

\subsection{{\it Chandra}}\label{data}

The data for this study were obtained from the {\it Chandra} Archive
and the ChASeM33 project.  The data included in this study as well as
the initial data processing steps and source detection are discussed
in detail in \citet{plucinsky2008}.  We produced light curves for all
of the sources in both the \citet{grimm2005} and \citet{plucinsky2008}
catalogs using the software package {\tt ACIS Extract}
\citep{broos2002}.  This software package is optimized to perform
measurements of sources observed with AXAF CCD Imaging Spectrometer
(ACIS) multiple times with different off-axis angles, and different
detector chips.  The output provides photon fluxes and errors from
each exposure as well as upper-limits for non-detections and a
spectrum for each source that combines the multiple exposures, taking
into account the different sensitivities and point spread functions of
different locations on the array. Furthermore, the output provides
short-term variability information by running a Kolmogorov-Smirnov
(K-S) test against a constant photon arrival rate during the brightest
detection.

The long-term light curves of all of the sources were calculated from
the {\tt ACIS Extract} output starting with the number of counts
within an aperture that enclosed 90\% of the energy in the point
spread function at the source's location on the detector in each
individual observation.  The local background was calculated from a
surrounding source-free area of the detector large enough to capture
50 counts. The background counts were then scaled to the appropriate
area and subtracted from the total counts.  The net counts were
converted to photon flux using the exposure maps created by the
ChASeM33 team \citep{plucinsky2008}.

This photon flux from each observation was then converted to a
luminosity estimate using a conversion factor of $2.7\times 10^{41}$
erg cm$^{2}$ photons$^{-1}$, which corresponds to the unabsorbed
0.35--8 keV luminosity of a source with photon index $\Gamma \sim 1.7$
and N$_H$=10$^{21}$ cm$^{-2}$.  Such spectral parameters are typical
for X-ray transient sources \citep{williams2006}.  The value chosen
did not affect our results, because our selection criteria were
determined based on relative fluxes, not absolute luminosities.  We
adopt this luminosity conversion to make the criteria that we used
more meaningful to the reader.

\begin{figure}
\centerline{\epsfig{file=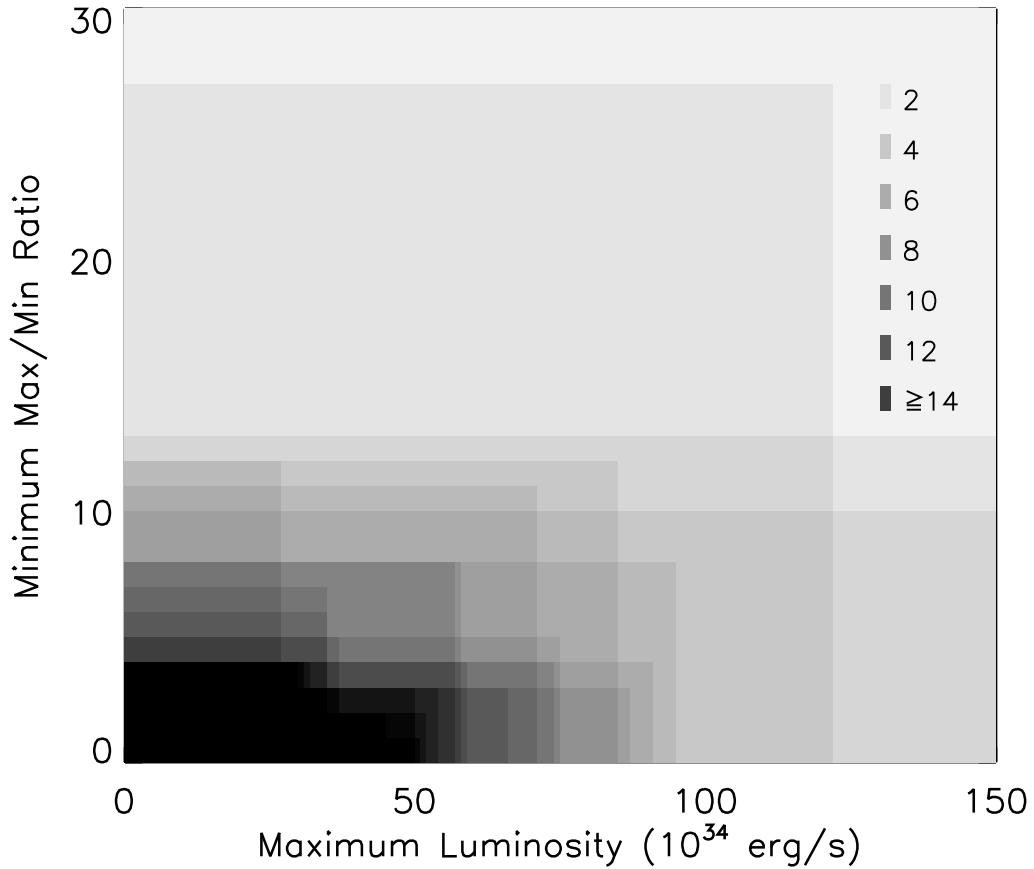,width=6.0in,angle=0}}
\caption{Grayscale plot of the effects of changing the maximum
luminosity and flux ratio ($h/l$) selection criteria for finding
transient candidates (see discussion in \S~\ref{data}).  Dark areas
denote larger numbers of sources passing the criteria.  The key in the
upper-right of the plot gives the number of sources represented by
some of the gray tones.  The sharp increase in number of sources
passing the search criteria when the maximum observed luminosity
criterion is set below about 4$\times$10$^{35}$ erg s$^{-1}$ and the
$h/l$ criterion is set below about 8 helped us determine the criteria
to use for initially flagging sources for follow-up analysis.}
\label{hists}
\end{figure}

We selected transient sources based on two criteria: first, the
maximum luminosity, which corresponds to the luminosity the source was
required to have in at least one observation; second, the minimum
luminosity ratio, which corresponds to the ratio $h/l$, where $h$ is
the maximum luminosity measured for a source in a single observation
and $l$ is the upper-limit on the faintest luminosity measured at the
location of the source in a single observation.  This ratio was
required to be larger than the adopted value for a source to be
considered as a transient candidate. A grayscale plot of the number of
sources passing our selection criteria as a function of the values of
those criteria is shown in Figure~\ref{hists}, where darker gray
corresponds to more sources passing the transient criteria.  The
number of transient candidates rises steeply in the region centered at
a maximum unabsorbed 0.35-8 keV luminosity cutoff of $\sim$4$\times
10^{35}$ erg s$^{-1}$ and a $h/l$ cutoff of $\sim$8.  We therefore
adopted these values to create our initial candidate list.
 
Based on these results, we flagged point sources that were (1)
detected in at least one observation with an unabsorbed 0.35-8 keV
luminosity of $>$4$\times 10^{35}$ erg s$^{-1}$ (1.5$\times 10^{-6}$
ph cm$^{-2}$ s$^{-1}$) at $\geq$4$\sigma$ significance, (2) not
detected at 1.5$\sigma$ significance in another observation, and (3)
had an upper-limit that was at least a factor of 8 fainter than the
maximum luminosity.  Errors were taken to be the standard deviation of
the number of counts, making the definition of a non-detection more
conservative than under the Gehrels approximation \citep{gehrels1986},
which yields larger errors and decreases the significance of
detections.  This selection yielded 8 initial candidates.

\begin{figure}
\centerline{\epsfig{file=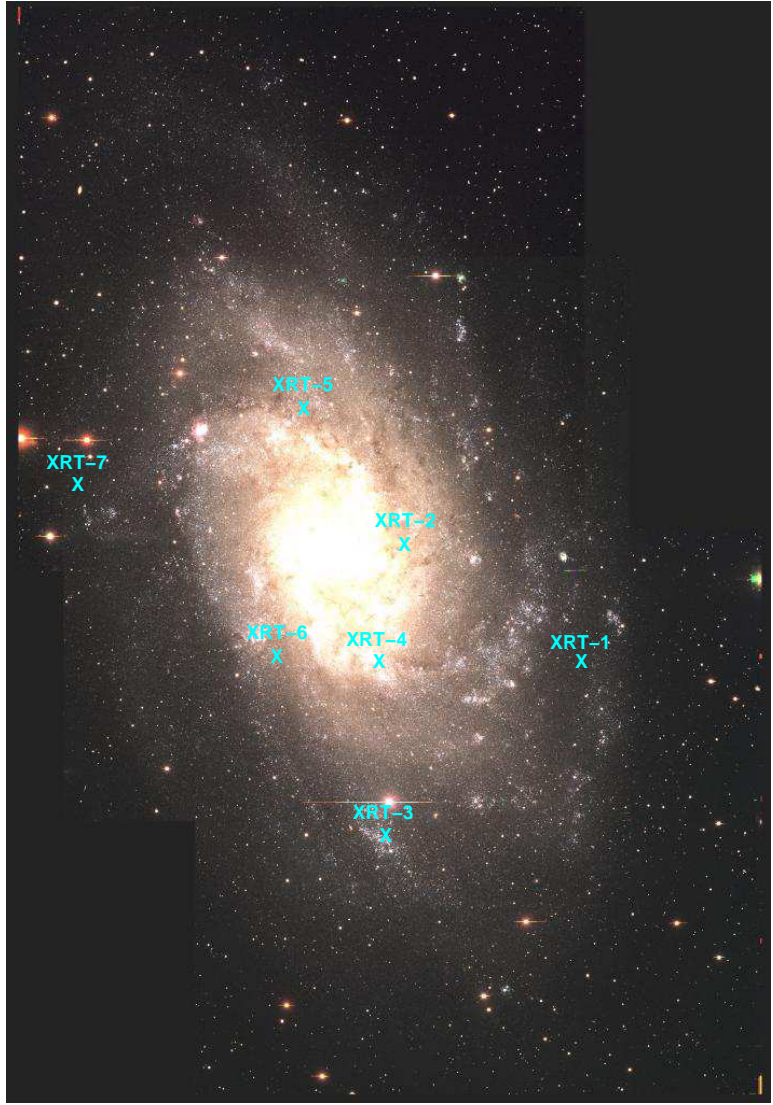,width=4.0in,angle=0}}
\caption{The locations of our 7 X-ray transient candidates are shown
on the Local Group Galaxy Survey image of M33 \citep{massey2006}.}
\label{loc}
\end{figure}

\begin{figure}
\centerline{\epsfig{file=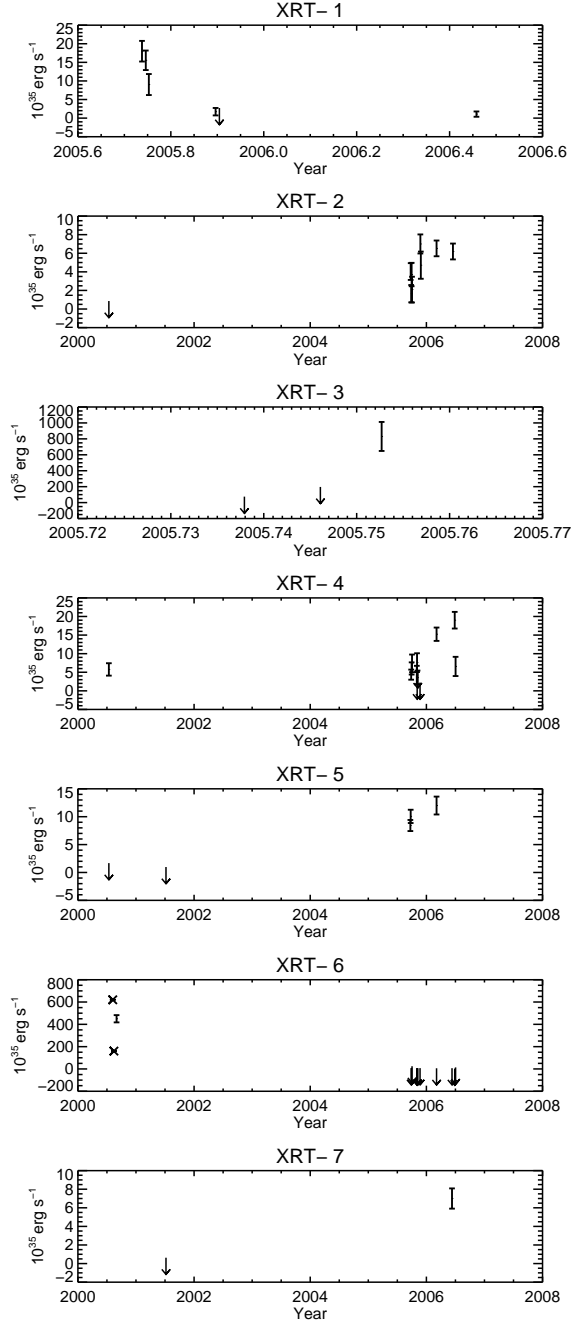,height=7.0in,angle=0}}
\caption{X-ray light curves for the 7 transient candidates.
Luminosities measured from archival {\sl XMM-Newton} data are marked with the
{\it X} symbol.  Error bars show standard deviation calculated from
photon statistics only and are appropriate for comparing the relative
measured fluxes for each source.  Absolute luminosities assume the
chosen spectral model detailed in Table~\ref{table} is correct and
therefore have much larger uncertainties.  Each source shows an
outburst that is more than 8 times brighter than its faintest
upper-limit.}
\label{lightcurves}
\end{figure}

Each observation of these candidates was then scrutinized by eye to look
for photometry problems as well as a visible detection and
non-detection.  One of the initial candidates had contamination from a
neighboring source and was therefore removed from the sample.  This
final culling produced 7 transient candidates.  The positions of these
sources are shown in Figure~\ref{loc}, and the light curves are shown
in Figure~\ref{lightcurves}. Images taken when these transient candidates were
detected at 4$\sigma$ significance are shown beside images taken when
the transient candidates were not detected in Figure~\ref{acis_im}.
Table~\ref{table} lists their observed properties.

Two of the seven candidates were previously cataloged by {\sl Chandra}
or {\sl XMM-Newton}.  These include XRT-2, which is source 145 in the
catalog of \citet{misanovic2006}, and XRT-6, which is the
previously-known supersoft transient source 207 in the catalog of
\citet{misanovic2006}. This latter source is also identified as source
J013409.9+303219 in the catalog of \citet{grimm2007}. The fact that
our search recovered \citet{misanovic2006} \#207 independently
suggests that our technique returns robust X-ray transient candidates.

We checked the literature for potential optical counterparts for our
transient candidates. We compared the ChASeM33 positions of X-ray
sources with known optical counterparts \citep[taken
from][]{hatzidimitriou2006,misanovic2006,shporer2006} to their
coordinates in the \citet{massey2006} catalog. The resulting average
shift between ChASeM33 and \citet{massey2006} coordinates was $\Delta
R.A. = -0.45 \pm 0.38''$ and $\Delta Dec.  = -0.21 \pm 0.39''$, which
is insignificant and consistent (within the errors) with the typical
ChASeM33 catalog positional error of $0.3''$. We therefore assume no
systematic offset in our search for counterparts, but we did include
an extra error component of 0.5$''$ (added in quadrature to our X-ray
position errors in Table~\ref{table}) to account for errors in
alignment between optical and X-ray positions.  Since this component
of the error is specific to the X-ray alignment with the
\citet{massey2006} data, it is only included in comparisons with this
dataset, not in the {\sl Chandra} position errors provided in
Table~\ref{table}.

\subsection{{\sl XMM-Newton}}

We analyzed the relevant portions of the {\sl XMM-Newton} survey of
M33 obtained from 2000 to 2003 to search for recurrent outbursts of
our transient candidates and to help constrain the spectral properties
of XRT-6. The individual {\sl XMM-Newton} observations were shorter by
about a factor of 10 than the {\sl Chandra} observations and therefore
the {\sl Chandra} data provided tighter constraints on the upper
limits for the transient sources.  All measurements were obtained
using the archival {\it XMM-Newton} data processed as detailed in
\citet{misanovic2006}.  Upper-limits were taken as the flux necessary
for a source with a spectrum of $\Gamma$=1.7 (assuming a column
density of $N_H=6\times 10^{20}$ cm$^{-2}$) to provide a detection
$\geq$3$\sigma$ above the background.

In two cases, {\sl XMM-Newton} data were not of sufficient spatial
resolution to obtain reliable upper-limits for the transient candidate
when the corresponding nearby source was active (XRT-3 and XRT-4).
Our inspections of the {\sl XMM-Newton} data showed no evidence for
X-ray sources at the locations of XRT-1, XRT-5, or XRT-7, the best
upper-limits for these candidates in any {\sl XMM-Newton} observation
were $\sim$3$\times 10^{35}$ erg s$^{-1}$ ($\sim$10$^{-6}$ ph
cm$^{-2}$ s$^{-1}$) for an energy range of 0.2-4.5 keV.  In addition,
even though XRT-2 was detected by {\it XMM-Newton}
\citep{misanovic2006} with a flux of 2.8 $\times$ 10$^{-15}$ erg
cm$^{-2}$ s$^{-1}$, the source could be transient.  The non-detection
of this source comes from an archival {\it Chandra} observation (OBSID
1730), which was taken 2000-Jul-12.  The detections of
\citet{misanovic2006} come from {\it XMM-Newton} observations of their
field 15 on 2002-Jan-27 (see \citealp{misanovic2006}, Table 1).
Therefore, while we were able to determine that only XRT-2 and XRT-6
were active during any of the {\sl XMM-Newton} observations of M33.
XRT-2 was detected in outburst by {\sl XMM-Newton} about 3 years
before the ChASeM33 detection indicating a recurrent nature of the
source.

\begin{figure}
\centerline{\epsfig{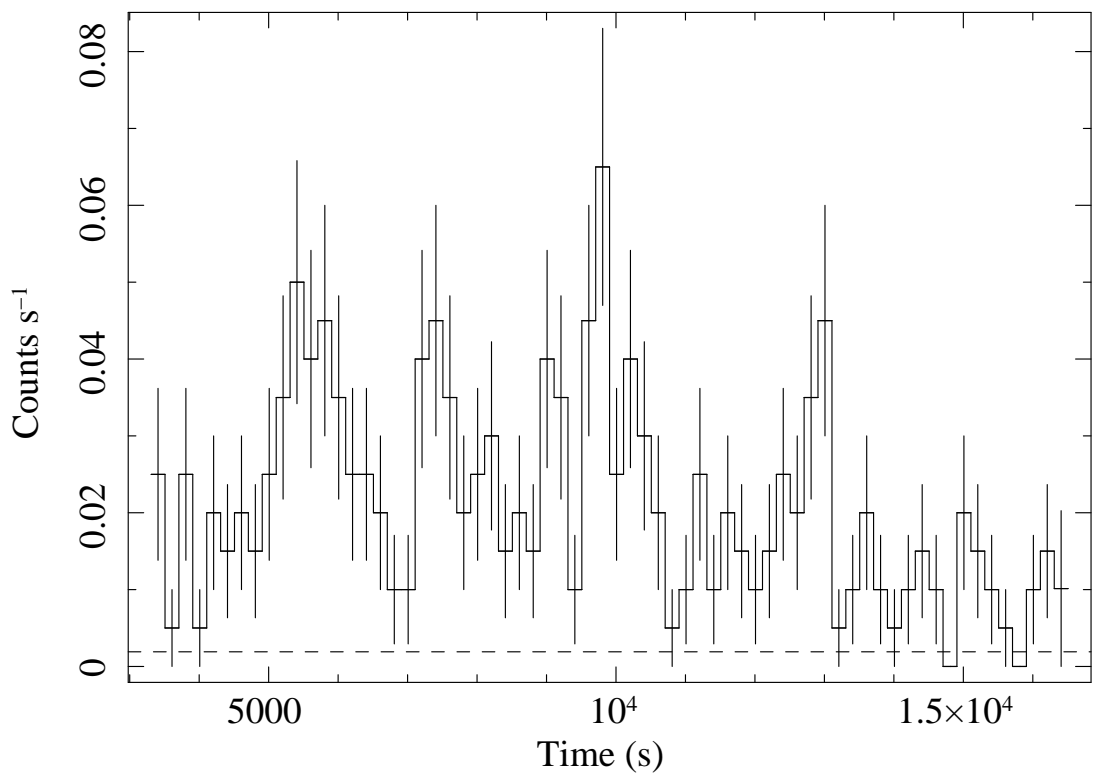}}
\caption{The {\sl XMM-Newton} EPIC-pn light curve of XRT-6 during the
2000, Aug. 2 observation, showing several Short outbursts with time
scales of 1000-2000 s. Single-pixel events were extracted in the
0.2-0.6 keV band. The average background level is indicated by the
horizontal dashed line. Zero time refers to UT 06:56:39.}
\label{xmmlc}
\end{figure}

We also made use of the {\sl XMM-Newton} EPIC-pn data pertaining to
the SSS XRT-6 in order to better constrain its spectral and temporal
parameters. Spectra were extracted using single-pixel events which
dominate at the low energies of interest. Although the number of net
source counts is low ($\sim$220 counts between 0.2 and 0.6 keV on
2000-Aug-02 and $\sim$105 counts on 2000-Aug-04) their confinement to
a narrow energy band allows us to derive characteristic temperatures
from a simple absorbed blackbody model fit. Light curves in the
0.2-0.6 keV band were extracted from the EPIC-pn data. The light curve
binned to 200 s from 2000-Aug-02 (taken with thick optical blocking
filter) with about 50\% higher average count rate as compared to
2000-Aug-04 (medium filter) is shown in Figure~\ref{xmmlc}.

\subsection{{\sl Spitzer}}\label{irdata}

We searched for mid-infrared counterparts to the seven X-ray transient
candidates in archival Spitzer observations of M33.  The entire galaxy
was mapped with the IRAC instrument (Guaranteed Time Observation
program 5; R. Gehrz, PI) between 2004 and 2006, with 6 observations
targeting the inner 30\arcmin\, of M33.  The IRAC instrument
\citep{fazio2004} features four imagers centered at 3.6, 4.5, 5.8 and
8.0 $\micron$ (channels 1 through 4) with mean pixel scales of
approximately 1\farcs22.  Each sequence in the M33 observations
consisted of 483 frames per channel with a 3 point, 1/2-pixel dither
for each position.  The exposure time per pixel in each channel was 12
s.  A preliminary analysis of data from this program is presented by
\citet{block2007}.

We utilized the 16 August 2004 IRAC observation of M33 (Astronomical
Observation Request ID 3638784) to search for mid-infrared (mid-IR)
counterparts.  The data used were the downloaded image mosaics from
the S14.0 version of the SSC pipeline.  We performed a visual search
for mid-IR counterparts to the seven X-ray transient candidates by
overlaying the positions of the X-ray sources onto the four IRAC
mosaics.  We found mid-IR counterparts to four of the transient
candidates (Table~\ref{spitztab}), though only one source (XRT-7) was
detected in all four IRAC channels.  We used SExtractor
\citep{bertin1996} version 2.3b2 to detect sources and perform
photometry on the images.  Stellar crowding in the outskirts of M33,
where the seven X-ray transient candidates are located, is not severe
in the IRAC images.  We found that most of the default parameters for
SExtractor were sufficient for detecting most of the point sources in
the X-ray transient candidate fields.  The parameters we adjusted were
the minimum number of pixels needed for detection (DETECT\_MINAREA =
3), the size of the region used to compute the background (BACK\_SIZE
= 32) and the background filtering size (BACK\_FILTERSIZE = 3).  In
addition, we specified a local computation of backgrounds for each
source (BACKPHOTO\_TYPE = LOCAL).

The input photometric parameters for SExtractor included the
instrumental gain for each channel, and the magnitude zero points as
defined in the Vega system by \citet{reach2005} and the IRAC version 3
data handbook.  We used the isophotal magnitudes (MAGAUTO) output by
SExtractor, computed from an aperture 5 pixels in radius and a
background annulus 24 pixels thick.  We then applied the aperture
correction factor for each channel documented in the IRAC data
handbook to obtain the final magnitude estimates.

\section{Results}

\subsection{Spectra}

The results of our {\tt ACIS Extract} total flux measurements are
supplied in Table~\ref{table} in the three ChASeM33 bands
\citep{plucinsky2008}: 0.35-1.1 keV (soft), 1.1-2.6 keV (medium), and
2.6-8 keV (hard).  These fluxes are the weighted mean fluxes of all
observations of each source.  The SSS XRT-6 had no flux in the hard or
medium band and is poorly fit by an absorbed power-law model.  The
best-fit photon index $\Gamma$=9.5 reflects the supersoft nature of
the source.  XRT-3 had no flux in the hard band
($\leq$2.5$\times$10$^{-7}$ photons cm$^{-2}$ s$^{-1}$) and was poorly
fit by a power-law model.  XRT-2, XRT-5, and XRT-7 exhibited flux in
all bands and have equivalent soft $\Gamma$ values.  XRT-1 and XRT-4
appear to be yet another spectral class, with most of their flux in
the hard band.  Similarly, their $\Gamma$ values are hard, with values
typical of those seen in HMXBs \citep[e.g.,][]{haberl2004}.
 
\begin{landscape}
\begin{figure}
\centerline{\epsfig{file=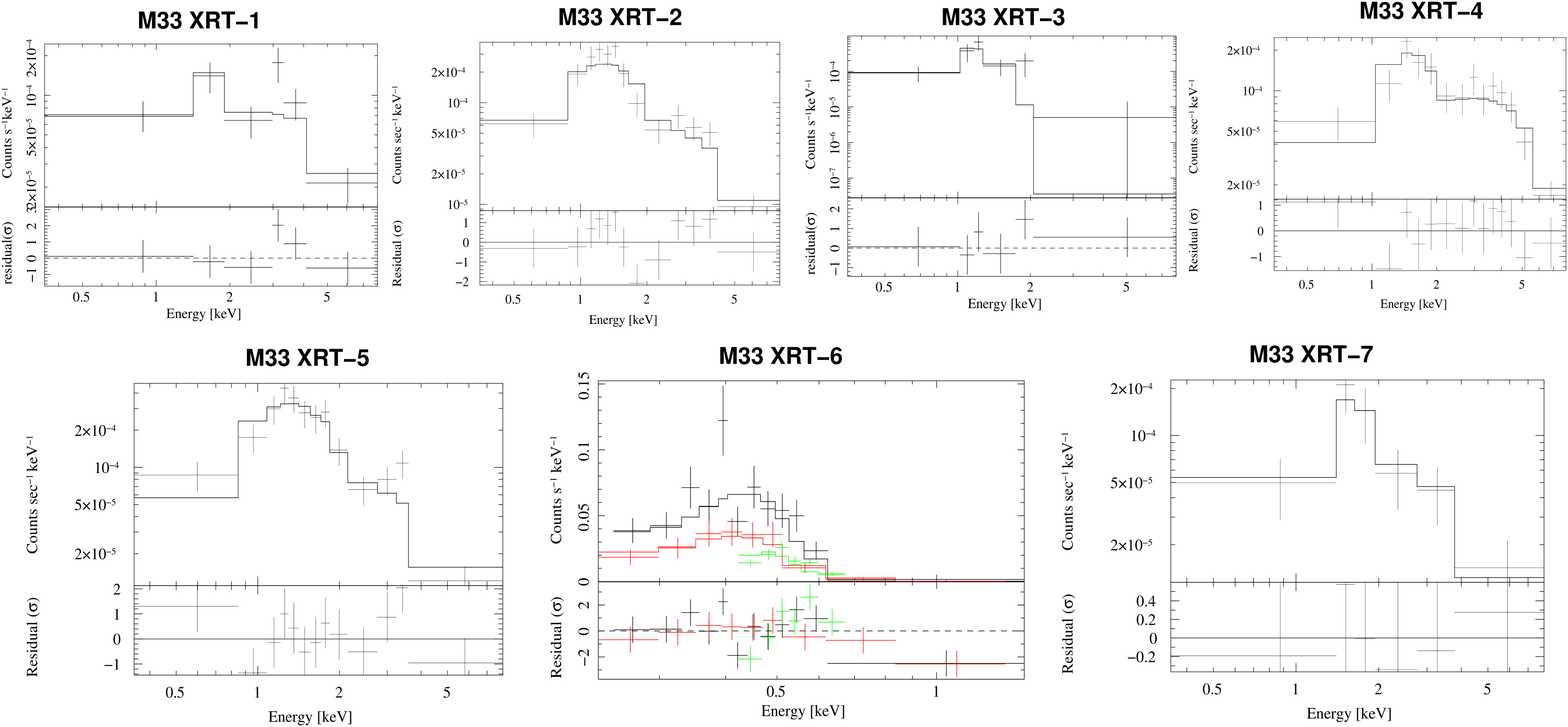,width=9.0in,angle=0}}
\caption{Spectral fits to the {\sl Chandra} data for each transient candidate.
XRT-1 and XRT-4 are significantly harder than the other
candidates. Absorbed power-law model fits are shown except for XRT-3
and XRT-6, where absorbed blackbody model fits are shown.  For XRT-6
(the SSS), the combined {\sl Chandra} (green) and {\sl XMM-Newton}-pn from
August 2 (black) and August 4 (red) spectral fit is shown.}
\label{spec}
\end{figure}
\end{landscape}

Fits to the spectra of the transient candidates were performed with
XSPEC 12.0 (Figure~\ref{spec}), initially assuming a power-law with
absorption.  For the candidates with good power-law fits ($\chi^2/\nu
< 1.5$; XRT-2, XRT-4, XRT-5, and XRT-7), we adopted the power-law as
the correct spectral model.  Otherwise, we attempted to find
better-fitting models.  If our data did not constrain the errors of a
parameter to better than a factor of 2, we do not quote errors for
that parameter.

Since the power-law fits to the spectra of XRT-1, XRT-3, and XRT-6,
had large $\chi^2/\nu$ values, we fit the spectra with some other
models to gain additional insight into their nature. The spectrum
extracted for XRT-1 was difficult to fit because of its hardness and
strong emission at $\sim$3 keV (see Figure~\ref{spec}). Because of
this data point, fits to the spectrum with thermal models were no
better than the original power-law fit.  We therefore accepted the
power-law model parameters when estimating the source peak luminosity
given in Table~\ref{table}.

The extracted spectrum of XRT-3 was best fit by either an absorbed
disk blackbody or an absorbed blackbody model, suggesting that this
transient candidate was in the high (thermal) state during the high
flux observation.  Since the disk blackbody and single temperature
blackbody fit equally well with similar temperatures, we provide the
parameters of the simpler black body model in Table~\ref{table}
($\chi^2/\nu = 0.8$ with $kT = 0.1$ keV and $N_H = 10^{22}$
cm$^{-2}$).  Our data are insufficient to distinguish between these
models.  Because of the low number of counts, the errors on the
spectral parameters were not better than a factor of 2. We note the
large range of absorption and temperature values allowed by the fit
($0<N_H<3 \times 10^{22}$ cm$^{-2}$ and $0.04<kT<0.5$ keV).

\begin{landscape}
\begin{figure}
\centerline{\epsfig{file=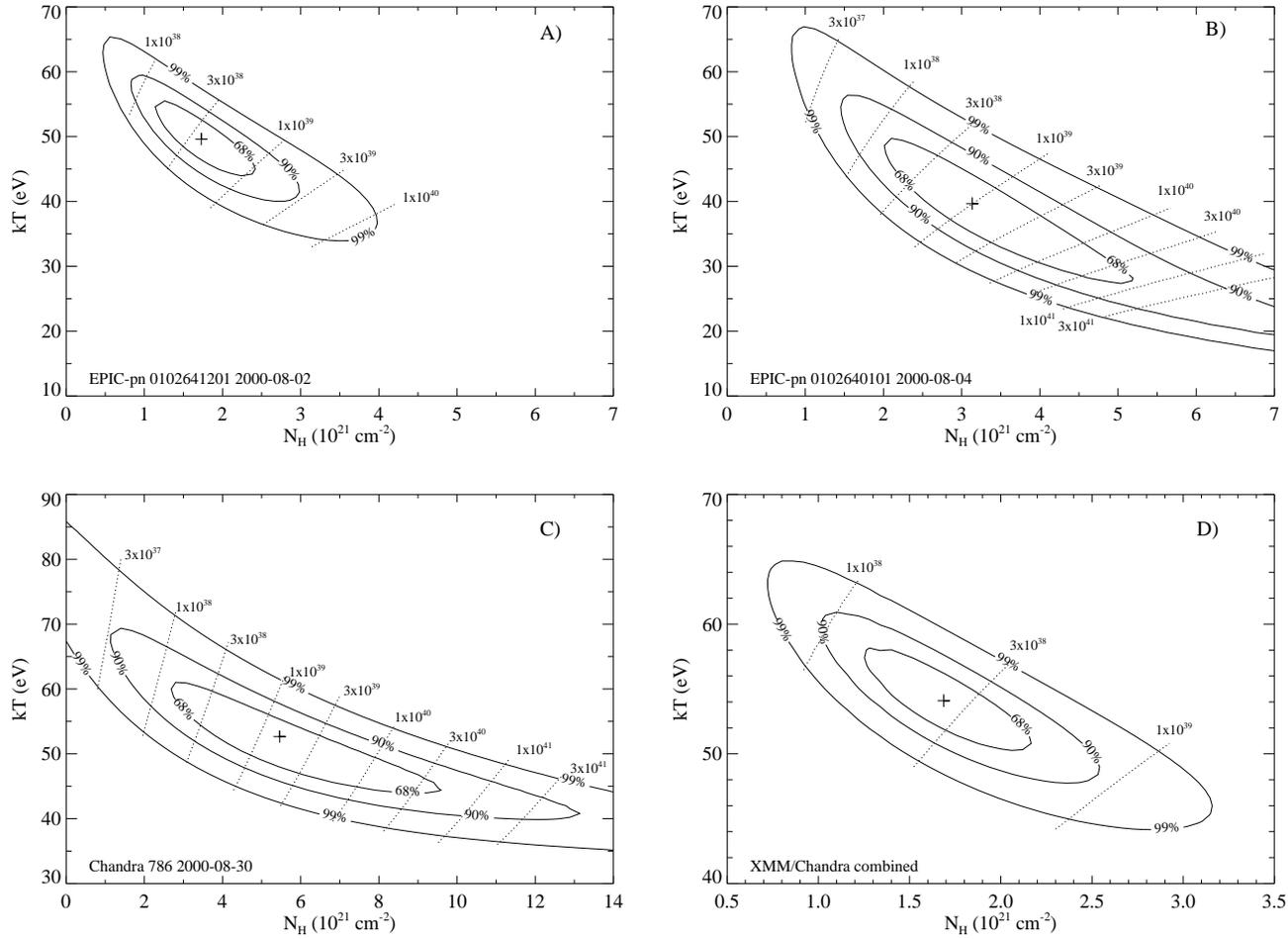,width=7.0in,angle=0}}
\caption{Column density - temperature confidence contours inferred
from fits to the three spectra of XRT-6 using an absorbed blackbody
model. The formal best fit parameter values are indicated by a
cross. {\it A-B:} The fitting results for the individual {\sl
XMM-Newton} observations. Lines of constant 0.2-2.4 keV luminosity (in
erg s$^{-1}$) are drawn on each plot, indicating the
absorption-corrected luminosity of the source with those spectral
parameters.{\it C:} The fitting results for {\sl Chandra}
observation. Lines of constant absorption-corrected 0.3-2.4 keV
luminosity (in erg s$^{-1}$) are drawn on each plot. {\it D:} The
results of a simultaneous fit to all of the spectra. Lines of constant
absorption-corrected 0.2-2.4 keV luminosity (in erg s$^{-1}$) are
drawn on each plot.}
\label{contours}
\end{figure}
\end{landscape}

Archival X-ray spectra for XRT-6 were extracted from both {\sl
XMM-Newton} and {\it Chandra} observations.  The power-law model did
not fit the spectra at all; therefore, we fit them with a blackbody
model.  The spectra, though taken at different times, proved to be
similar, showing overlapping best-fit parameters for temperature and
absorption.  The observed 0.35-1.1 keV flux decreased from
1.2$\times$10$^{-13}$ erg cm$^{-2}$ s$^{-1}$ on 2000-Aug-2 to
2.9$\times$10$^{-14}$ erg cm$^{-2}$ s$^{-1}$ on 2000-Aug-4.  Applying
the same model to the {\it Chandra} spectrum from 2000-Aug-30 yields a
flux of 6.0$\times$10$^{-14}$ erg cm$^{-2}$ s$^{-1}$. Column density -
temperature confidence contours derived from the three spectra largely
overlap, as shown in Figure~\ref{contours}, and a combined fit
assuming constant temperature and column density yields $kT$ =
54$\pm$4 eV and $N_H$ = 1.7($\pm$0.5)$\times$10$^{21}$ cm$^{-2}$,
assuming elemental abundances from \citet{anders1989}.  Column
densities in excess of N$_{\rm H}\sim$1.9$\times$10$^{21}$ cm$^{-2}$
are excluded if we assume an intrinsic source luminosity (0.2--2.4
keV, distance 795 kpc) for the 2000-Aug-02 observation below the
Eddington limit \citep[see also][]{pietsch2007}. This absorption value
is lower than that measured with the {\it Chandra} data alone
\citep{grimm2007}, possibly due to the better sensitivity of {\it
XMM-Newton} at low energies or variable intrinsic absorption in the
source.

We note that large extrapolation uncertainties go into estimating the
bolometric luminosity of a very soft source, making the estimate quite
model-dependent \citep{heise1994}. With more detailed spectra and more
sophisticated modeling, the luminosity estimate for XRT-6 may
decrease; however, such a decrease would require a decrease in the
absorption correction. Therefore our measured upper-limit for the
absorption column is secure against such changes in the spectral
model. Furthermore, our absorption limit assumes the 0.2--2.4 keV
X-ray luminosity as a conservative lower limit on the bolometric
luminosity.  We have restricted the luminosity estimate to the energy
band that we actually observe, minimizing the model dependence of the
estimate. Since the observed band is only a small fraction of the
spectral energy distribution, our restriction on the column density is
conservative.  Better knowledge of the intrinsic source spectrum and
bolometric luminosity would further decrease the upper limit for the
column density.
 
We estimated the absorbed and unabsorbed 0.35--8 keV luminosities of
the transient candidates with {\tt XSPEC}, assuming the best-fit
spectral model parameters. The unabsorbed values are shown in the
light curves in Figure~\ref{lightcurves}.  Both values are provided for
the observation with the greatest flux in Table~\ref{table}.  The
photon fluxes given in the table were calculated before the spectra
were fit and come from the exposure maps created by the ChASeM33 team.
They are meant to provide a consistent way of measuring the ratio of
the highest to lowest observed flux and an important comparison
between the best-fitting spectral parameters and the hardness ratios.

\subsection{Timing}  

Although our temporal sampling is somewhat erratic, it reveals some
interesting characteristics of the transient candidates.  In addition,
the long-integration observations provide some sensitivity to short
outbursts while the sources were active.

Our sampling covers the decay of M33 XRT-1 fairly well, as can be seen
in the light curves of Figure~\ref{lightcurves}.  This source decayed
by nearly a factor of 2 in just 4 days, suggesting an $e$-folding
decay time of $\sim$1 week.  As the source was not detected at the
1.5$\sigma$ level in the next observation, 58 days after the brightest
detection, we can conservatively place a 58 day upper limit on the
decay time of the source, which is typical for a $Be$ HMXB \citep[see
examples in][]{laylock2003}.  The decay time of XRT-3 could not be
constrained in a meaningful way from our data, but the rise time of
XRT-3 appears to be rapid, increasing in flux by a factor of 7 in
under 3 days and a factor of at least 11 in less than 6 days.  Such a
rise time is similar to that of Galactic LMXB transients and black
hole binaries \citep[e.g.,][]{chen1997}.

The long-term lightcurve for XRT-4 does not provide clear evidence
that it is a true X-ray transient.  It appears to be a faint,
persistent source most of the time.  In 2005, both of our
non-detections have 1$\sigma$ upper-limits near the bottom of the
error bars of the detections, suggesting that the source was at a
similar low flux during the shorter observations when it was not
detected.  There are no detections of the source in the {\it
XMM-Newton} archival data, but the most sensitive observation at this
location (observation 12a in \citealp{misanovic2006}), taken
2000-Aug-02 provides an 0.2-4.5 keV absorbed luminosity upper-limit of
6$\times$10$^{35}$ erg s$^{-1}$, which is similar to the lower
detected luminosities in the {\it Chandra} data.  On the other hand,
the source underwent an outburst in late June of 2006.  It then
decayed by a factor of 3 from 2006-Jun-26 to 2006-Jul-01.
Furthermore, this source shows short-term variability, has a hard
spectrum, and is associated with a blue optical counterpart candidate,
making it a possible HMXB transient X-ray source.

The candidates XRT-5 and XRT-7 do not have enough temporal coverage to
sample rise or decay times.  For these candidates, the fact that they
were never detected by {\sl XMM-Newton} offers some supporting
evidence that these sources are transients, but again, the sensitivity
of the {\sl XMM-Newton} observations provides 0.2-4.5 keV absorbed
luminosity upper limits of 5$\times$10$^{35}$ erg s$^{-1}$, which is
only slightly below the brightest detected luminosity in the {\it
Chandra} data.

The {\sl XMM-Newton} data for XRT-6 yield a short decay time, as the
flux dropped by a factor of 4 in the 2 days between observations.
These measurements suggest an $e$-folding decay time of $\sim$1.3
days.  However, the lightcurve of this transient candidate is clearly more
complex than a simple exponential decay, as demonstrated by its flux
in the {\sl Chandra} observation 26 days after the {\sl XMM-Newton}
observations and by its short-term variability.

\begin{figure}
\centerline{\epsfig{file=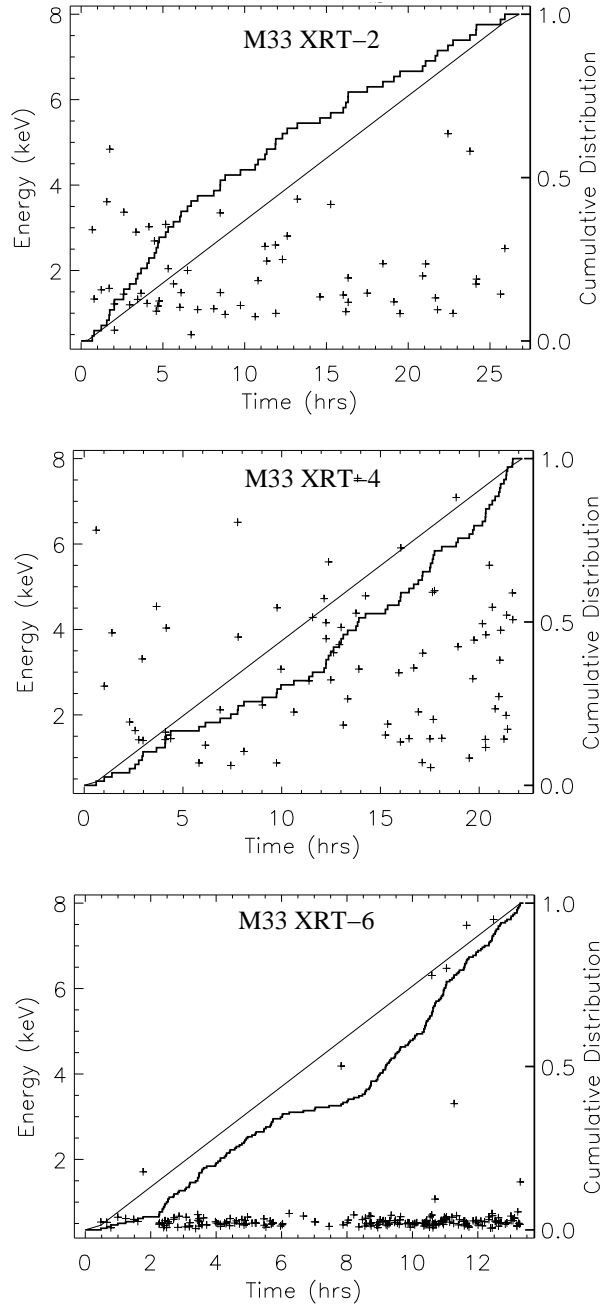,height=7.0in,angle=0}}
\caption{Unbinned photon arrival times for XRT-2, XRT-4 and XRT-6
  during Observations 6376 (top), 6387 (middle) and 786 (bottom),
  respectively.  Black points mark the time and energy of each
  detected photon.  The histogram shows the cumulative fraction of
  detected photons, and the line marks the cumulative fraction
  expected for a source with constant flux.}
\label{lc}
\end{figure}

The K-S tests for short-term variability (i.e. flares or bursts within
a single observation) of XRT-1, XRT-3, XRT-5, and XRT-7 gave
probabilities of 0.1 or higher that the source was constant during the
observation with the most counts.  On the other hand, the sources
XRT-2, XRT-4 and XRT-6 each showed evidence for short term variability
while they were active.  Our K-S test provides probabilities of only
2.2$\times$10$^{-2}$, 4.4$\times$10$^{-3}$ and 6$\times$10$^{-9}$ that
XRT-2, XRT-4 and XRT-6 had constant flux during ObsIDs 6376, 6387, and
786, respectively.  The photon arrival times from those ObsIDs are
shown in Figure~\ref{lc}.  While the distribution for XRT-2 hints at a
higher flux early in the observation and XRT-4 exhibits a higher flux
later in the observation, the most dramatic variability observed is
that of XRT-6, which shows two distinct bursts during the observation.
One burst lasted for $\sim$4 hours, beginning 2 hours into the
observation.  After a 2 hour break, there was another burst that
continued to the end of the observation.  Furthermore, in the {\sl
XMM-Newton} data from the brightest detection (2000-Aug-02), the
source exhibits strong variability with outbursts on timescales of
1000-2000~s.  Similar variability was observed during the last third
of the observation on 2000-Aug-04.

\subsection{Optical and Infrared Counterparts}

We searched several available data archives and published catalogs for
counterparts for the transient X-ray sources at optical and infrared
wavelengths.  These included the Local Group
Survey\footnote{http://www.lowell.edu/$\sim$massey/lgsurvey/} \citep{massey2006},
the CFHT M33
survey\footnote{http://www.astro.livjm.ac.uk/$\sim$dfb/M33/} \citep{hartman2006},
the DIRECT
project\footnote{http://cfa-www.harvard.edu/$\sim$kstanek/DIRECT/} \citep{macri2001},
and the Wise Observatory M33
survey\footnote{http://wise-obs.tau.ac.il/$\sim$shporer/m33/} \citep{shporer2006}
catalogs, as well as imaging from the {\it Spitzer} data archive (see
section \S~\ref{irdata}).

\begin{landscape}
\begin{figure}
\centerline{\epsfig{file=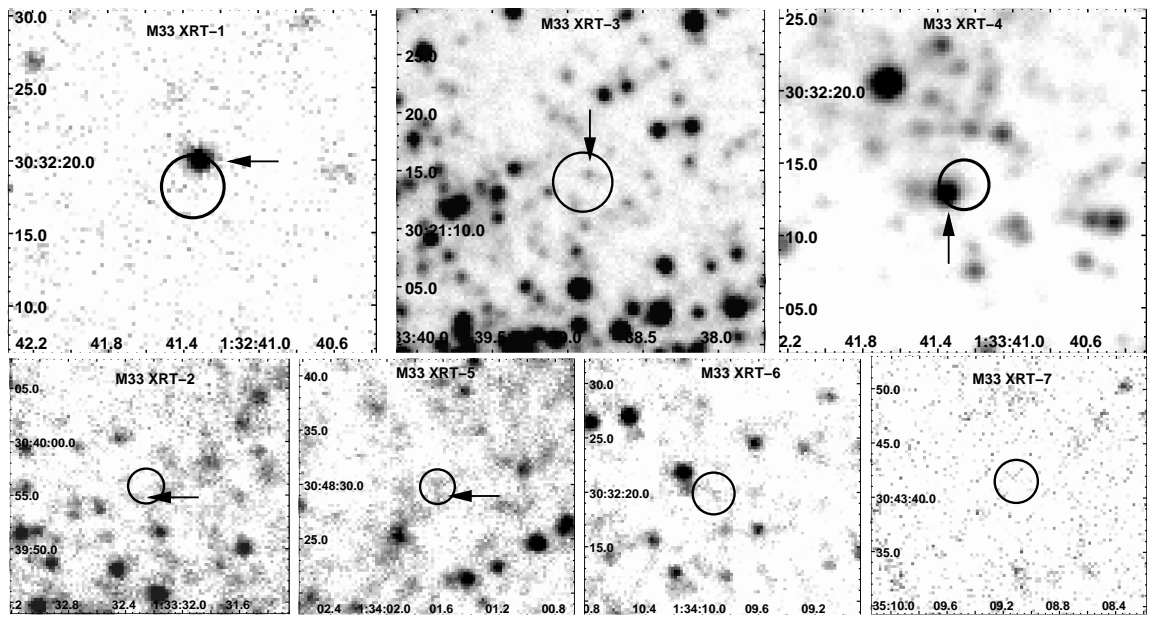,width=6.0in,angle=0}}
\caption{$B$-band images of the locations of the 7 transient
candidates taken from the Local Group Survey
\citep{massey2006}. Circles are 3$\sigma$ in radius (after adding
0.5$''$ in quadrature to the X-ray position errors to account for
X-ray - optical alignment). The best counterpart candidates from the
\citet{massey2006} data are indicated with arrows.  Bright blue stars
overlap the XRT-1 and XRT-4 circles.  A bright red variable star
(faint in the $B$-band image) is just outside the east side of the
XRT-3 circle.  Faint stars are cataloged just inside the error circles
of XRT-2 and XRT-5.  No clear counterpart candidates were found for
XRT-6 or XRT-7.}
\label{b_ims}
\end{figure}
\end{landscape}

Several of the new potential X-ray transients have good optical
counterpart candidates (Figure~\ref{b_ims}).  This fact suggests that
the sample contains several HMXBs. Galactic transient LMXBs typically
have $-2<\rm{M}_V<5$ ($22.5<\rm{m}_V<29.5$ in M33) {\it in outburst}
and are fainter in quiescence \citep{vanparadijs1994}, making LMXBs in
M33 difficult to detect in ground-based imaging.  Nevertheless, both
XRT-1 and XRT-4 are within 2$''$ of a bright blue star in the
\citet{massey2006} M33 catalog (IDs J013241.31+303220.0 and
J013341.33+303212.6), and three of the other transient candidates have
blue or variable stars within their 3$\sigma$ error circles.

There are also transients that have good mid-infrared counterpart
candidates.  At these wavelengths, light from the donor star dominates
the emission \citep[e.g.,][]{shahbaz1996}.  We therefore expect these
systems to look like single stars, as is the case with all but one of
the counterpart candidates.

\subsubsection{Optical Counterparts}

\begin{figure}
\centerline{\epsfig{file=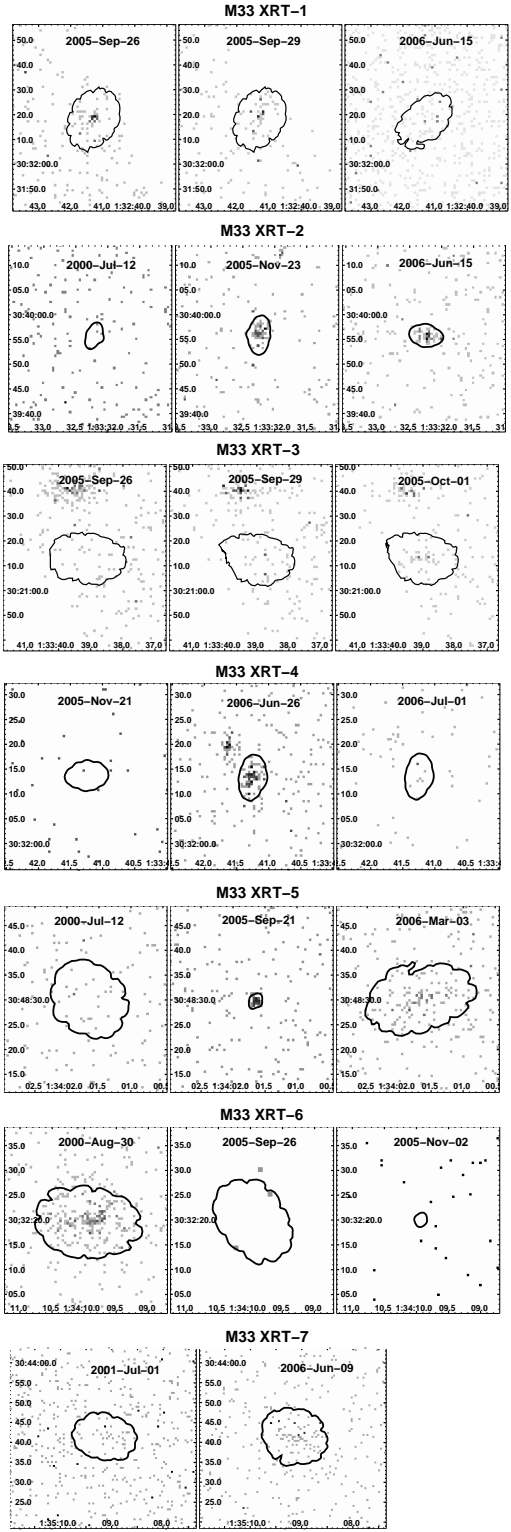,height=8.0in,angle=0}}
\caption{X-ray images of the 7 transient candidates at different times.
The dates were chosen to sample on and off times for each candidate.
The background levels are different for each observation due to the
different exposure times.}
\label{acis_im}
\end{figure}

The magnitude ($V=20.29$) and color ($B-V=0.03$) of the XRT-1
counterpart suggest that the star is a moderately extincted ($E_{B-V}
\sim 0.28$) late O or early B star in M33. This counterpart candidate
is not in a crowded region of M33, making the identification
surprisingly clear. Furthermore, this star is known to be variable
\citep[CFHT~250258, D33~J013341.3+303212.7;][]{hartman2006,macri2001}
Although the X-ray position is off the star by approximately 2$''$,
the X-ray position is not tightly constrained because the source is on
the detector edge.  The X-ray position is off to the southeast, and
the detection is on the northwest edge of the ACIS-I detector, as
shown in Figure~\ref{acis_im}.  Therefore any flux that could have
fallen farther to the northwest would have been off the detector edge.
The fact that the point spread function of the source spills over the
detector edge to the northwest increases the likelihood that the {\it
Chandra} position is skewed to the southeast.

The only cataloged star within the XRT-2 3$\sigma$ error circle
(1.6$''$) is J013332.34+303954.9 in the \citet{massey2006} catalog.
This is a faint blue star with $V=22.8$ and $B-V=0.1$.  With this
brightness and color, the star could be a moderately extincted early
$B$ star in M33.  If this star is the counterpart, XRT-2 would also be
a HMXB, a reasonable possibility since this transient is recurrent and
has a fairly hard spectrum ($\Gamma\sim 1.6$).

For XRT-3, the nearest optical counterpart is an uncataloged blue star
within the error circle.  This star is indicated with the arrow in
Figure~\ref{b_ims}.  We performed aperture photometry on this object
and a neighboring star in the \citet{massey2006} catalog using the
images from the Local Group Survey.  After correcting our magnitudes
so that our photometry for the cataloged star matched those of
\citet{massey2006}, the star has $V=23.3$ and $B-V=-0.1$, making it a
moderately extincted B star in M33.  Therefore XRT-3 may be another
HMXB transient; however, the soft X-ray spectrum is not consistent
with an HMXB transient.  On the other hand, another potential optical
counterpart is another known variable star (CFHT~320133).  Photometry
from the \citet{massey2006} catalog (ID J013339.13+302113.2,
$V=20.36$) shows that it is quite red ($B-V=2.00$).  The star is
redder than low-mass dwarfs, making the foreground reddening toward
M33 inadequate to explain its color. The star is therefore a good
candidate for an extincted red supergiant or red helium-burning star
in M33.  As is the case with XRT-1, this source is also near the
detector edge.  In this case it is near the southeast edge of the
detector (see Figure~\ref{acis_im}), increasing the likelihood that
some flux spilled off the southeast edge of the detector, skewing the
position to the northwest.  The X-ray position is misaligned with the
position of the star in this direction, increasing the plausibility
that this star is the true counterpart, but this candidate is less
certain than that of XRT-1 due to the higher density of stars in the
area and the blue candidate closer to the measured X-ray position.
From the density of variable stars with 1$'$ of XRT-3 in the
\citet{hartman2006} catalog (0.0095 arcsec$^{-2}$), the probability of
finding a random variable star within 2$''$ of XRT-3 is 0.12.

The XRT-4 counterpart candidate is likely a bit higher in mass; with
$V=18.13$ and $B-V=-0.18$, it is likely a late O supergiant.  In
addition, this counterpart candidate shows periodic variability with P
= 20.16 days \citep[W21058;][]{shporer2006}.  Although the
positional agreement of XRT-4 with the optical counterpart candidate
is also only 1.5$''$, the optical variability and the hardness of the
X-ray spectrum make this star a good counterpart candidate.

There is a variable optical counterpart candidate for XRT-5
\citep[CFHT~144719;][]{hartman2006} inside of the 3$\sigma$ error
circle, 1.1$''$ south of the X-ray position. This star was not
included in the \citet{massey2006} catalog, but has $\langle g \rangle
= 22.4$ and $\langle g-r \rangle =0.5$.  This brightness and color
allows the possibility that this star is in the Galactic halo, but the
bright mid-IR counterpart (see \S~3.3.2) suggests the presence of a
highly absorbed object.  If so, then XRT-5 could be a heavily
extincted HMXB in M33.

There are no clear optical counterpart candidates for XRT-6 or XRT-7.
XRT-6 appears to have some optical emission within its error circle,
but no stars have been cataloged at this position, suggesting the
counterpart is too crowded to be measured in ground-based data.  The
area surrounding XRT-7 shows no $B$-band emission at all in the
\citet{massey2006} images.

The optical counterparts provide some insight into the bright
transient population of M33.  Apparently, at least 2 (and possibly
five) of the transients (XRT-1, XRT-4, and possibly XRT-2, XRT-3
and/or XRT-5) are HMXBs.  This result demonstrates a difference
between M33 and the other earlier and more massive spirals of the
Local Group, M31 \citep{williams2006} and the Galaxy \citep{chen1997}.
The bright transient population of M33 is not dominated by LMXBs, and
in fact may contain a large fraction of HMXBs.  Since HMXBs are more
short-lived than LMXBs, this low LMXB transient fraction suggests a
more dominant young population, reflecting the later morphology of
M33.

\subsubsection{Infrared Counterparts} 

\begin{figure}
\centerline{\epsfig{file=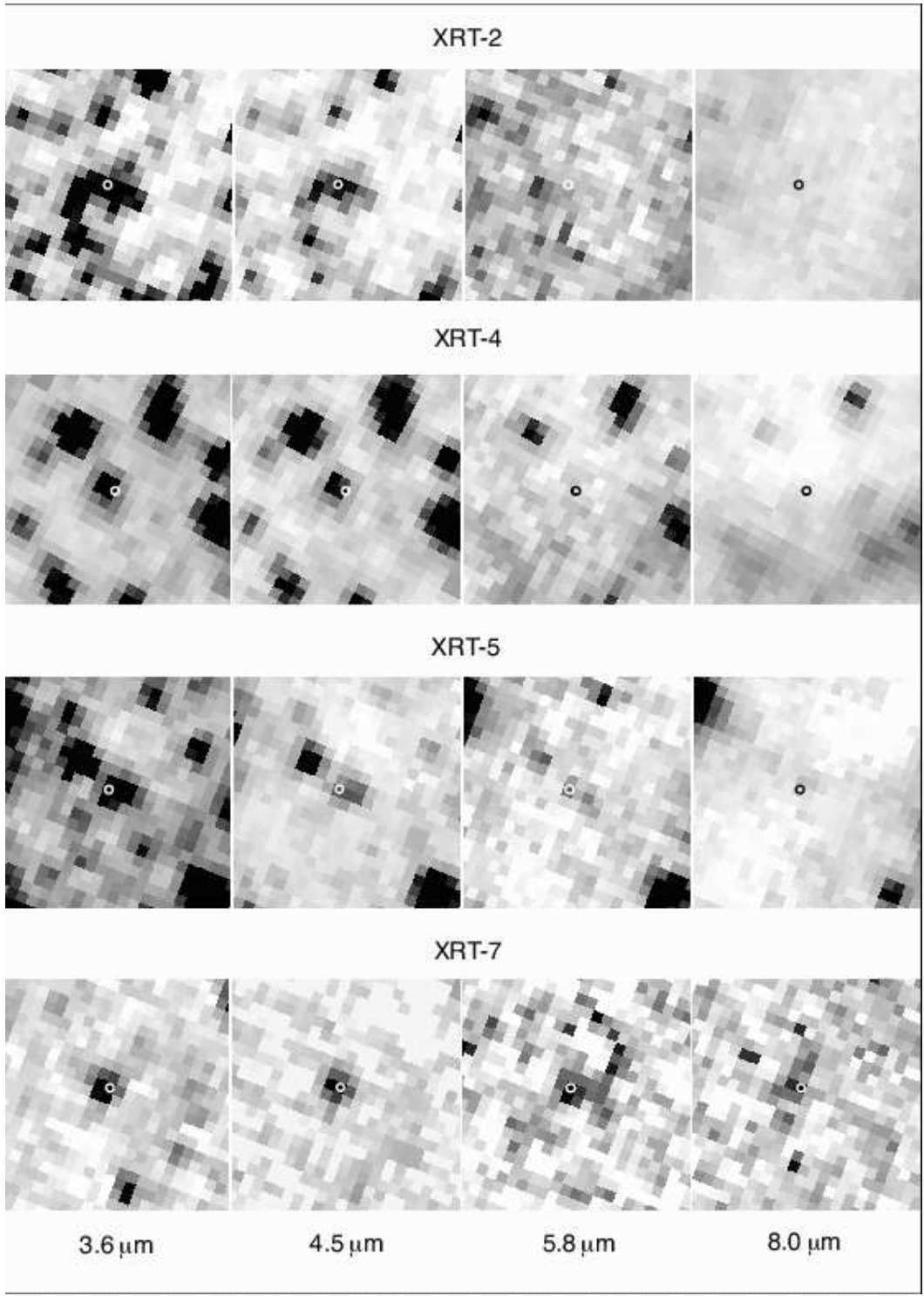,width=5.5in,angle=0}}
\caption{Mid-Infrared counterparts to the X-ray transient candidates
in M33.  The X-ray position of each source is marked on closeup images
of all four IRAC channels.  Each field is 25\arcsec\, square in size,
with north at the top and east to the left.  XRT-2 appears to be
coincident with an extended source.  XRT-2, XRT-4 and XRT-5 are
co-incident with sources with blue IR colors similar to those of
stars.  XRT-7 has IR colors that are consistent with broad-line AGNs.}
\label{spitzer}
\end{figure}

In Figure~\ref{spitzer} we present images of the X-ray transient
candidates showing counterparts in the IRAC data.  The positions of
the X-ray transient candidates are marked on each image.  Transient
candidates 1, 3, and 6 do not show obvious mid-IR emission and were
not detected by SExtractor.  Sources XRT-2, XRT-4 and XRT-5 were
detected only in the first two IRAC channels, while source XRT-7 is
detected in all four channels.

Using the photometric measurements (Table~\ref{spitztab}) we computed
the IRAC colors [3.6]\,-\,[4.5] and [5.8]\,-\,[8.0].  Three infrared
counterpart candidates (XRT-2, XRT-4, XRT-5) show colors that are
consistent with either constant or declining flux (F$_{\nu}$) with
increasing wavelength, though without detections in all four bands we
are unable to better determine whether these objects are stellar
sources within M33 or background galaxies from their fluxes alone.  On
the other hand, comparing the colors of XRT-7 with those of sources
detected in the IRAC shallow survey \citep{stern2005} we find that the
colors of this object ([3.6]\,-\,[4.5] = 1.1$\pm$0.6, [5.8] - [8.0] =
1.0$\pm$0.5) strongly suggest that it is an AGN.  The lack of optical
emission and high X-ray absorption from XRT-7 are also consistent with
an obscured AGN. This infrared counterpart therefore rules out the
possibility that another transient (XRT-7) is a LMXB, leaving only 3
of 7 transients with the potential of being LMXBs.

The sizes of the mid-IR counterparts can also help in the
determination of the object type.  Although the large pixels and PSF
of {\sl Spitzer} make blending an obvious problem, one of the infrared
counterparts was much larger than the {\sl Spitzer} PSF.  The
counterpart of XRT-2 was measured to have a full-width at half-maximum
of 7.9 pixels (9.5$''$) at 3.6 $\mu$m.  The size reduces to 6.5$''$ at
4.5 $\mu$m, which is still more than double the IRAC PSF
($\sim$2.5$''$).  The region is clearly very crowded in the {\sl
Spitzer} data, but of all of our counterparts, this one appears most
likely to be extended.  Such an extended mid-IR source could be
unresolved stars in M33, which would be consistent with the large
difference in size in different bandpasses, or a background galaxy.
However, the likelihood of a background galaxy is low considering the
X-ray variability and moderate absorption measured.  Therefore, we
suggest that this counterpart is a blend of stars in M33, still
consistent with the possibility that XRT-2 is an LMXB or an HMXB.

The combination of optical and IR properties of the counterparts can
also be effective for constraining object type.  XRT-4 is the only
transient with a strong counterpart candidate in both the optical and
mid-IR data, likely because XRT-4 has the brightest optical
counterpart candidate, which is probably a supergiant star.  Fainter
blue stars (such as the counterparts for XRT-1 and XRT-3) would
require deeper IR imaging to detect. The lack of a clear, bright
optical or IR candidate for XRT-3 is consistent with the possibility
that it is a LMXB.  On the other hand, the counterparts for XRT-2 and
XRT-5 are detected.  XRT-2 could be in a small cluster that is
extincted.  This would explain the bright and extended appearance in
the mid-IR, but the faint appearance in the optical.  XRT-5 has the
reddest variable optical counterpart, which may explain its appearance
in the mid-IR images. If the source is located in M33, it must be
heavily absorbed and have an IR excess to explain its bright mid-IR
fluxes as seen in the Spitzer data.  Such IR excesses are common in
$Be$ systems \citep{cote1987}, allowing the possibility that XRT-5 is
an HMXB, despite its relatively soft X-ray spectrum.

\section{Conclusions}

We have discovered 6 new candidate transient X-ray sources in M33; one
of which (XRT-2) is a recurring transient source.  A seventh
previously-known transient was also found by our search technique.
These sources vary by at least a factor of 8 in flux in the {\it
Chandra} observations considered here, and their spectra suggest they
belong to a few different classes.  For example, XRT-1 and XRT-4 are
hard sources similar to HMXB transients.  In contrast, XRT-2, XRT-5,
and XRT-7 have soft spectra that are well-fitted by a power-law model.
XRT-3 has an even softer spectrum similar to those of LMXB transients
in the high state, and XRT-6 is a known SSS.

Optical counterpart candidates for XRT-1, XRT-2, XRT-3, XRT-4, and
XRT-5 were identified from the DIRECT \citep{macri2001},
\citet{massey2006}, \citet{shporer2006}, and CFHT \citep{hartman2006}
catalogs.  The density of other faint stars near the locations of
XRT-2, XRT-3, and XRT-5 make their counterpart candidates less
reliable than the bright counterpart candidates of XRT-1 and XRT-4.
These two candidates are bright blue variable stars, strongly
suggesting that XRT-1 and XRT-4 are HMXBs.

Infrared counterparts of XRT-2, XRT-4, XRT-5, and XRT-7 were detected
in {\it Spitzer} images of M33.  The mid-IR color of XRT-7 suggests it
is an AGN, consistent with its highly absorbed X-ray spectrum and the
lack of an optical counterpart candidate.  The mid-IR counterpart for
XRT-2 has a color typical of stars, but it is extended, possibly
because of crowding.  The mid-IR counterpart of XRT-4 is consistent
with being the same bright star seen in the optical.  The mid-IR
counterpart of XRT-5 also has the color of a star, but it is bright
compared to the optical counterpart candidate, suggesting the optical
counterpart is heavily extincted.  If so, XRT-5 is another good HMXB
transient candidate.

Our multi-wavelength measurements allow the possibility that at most
three (XRT-2, XRT-3 and XRT-5) of these 6 sources could be LMXBs.
While these three transients are particularly difficult to classify,
only XRT-3 has no mid-IR counterpart and is much softer than expected
for an HMXB.  Therefore XRT-3 is the best LMXB candidate in our
sample.

Although our sparse sampling does not allow us to constrain the
transient rates or duty cycles, the source types and locations allow
for some interesting comparisons with the Galaxy and M31.  Apparently,
the transient population of M33 is not dominated by LMXBs.  This low
LMXB fraction is different from the bright X-ray transients in M31,
where most are found in the bulge \citep{williams2006} and have
optical counterparts consistent with LMXBs \citep{williams2006bh6}.
On the other hand, the M31 transients are similar to those of the
Galaxy.  For example, the distributions of transient peak luminosities
and decay times in M31 are very similar to those of the Galaxy
\citep{williams2006,chen1997}.  Our results now suggest that these
populations are somewhat different from that of M33, whose brightest
transients appear to be spread over the whole galaxy and to contain a
significant fraction of HMXBs.  M33 is known to be a later-type, more
intensely star-forming galaxy than M31 or the Galaxy.  These
morphological differences are reflected in the metallicities, stellar
populations, and apparently the bright X-ray transient populations of
the galaxies.

Finally, the potential of some of these sources for containing
stellar-mass black holes makes their discovery of particular interest
for studies of black hole formation and accretion.  The final catalog
of sources from the ChASeM33 data is currently under construction, and
our search for transients will be extended to include all of those
sources contained in the full survey data set when the final source
catalog is available.

Support for this work was provided by the National Aeronautics and
Space Administration through {\sl Chandra} Award Numbers G06-7073A and
GO6-7073B issued by the {\sl Chandra} X-ray Observatory Center, which
is operated by the Smithsonian Astrophysical Observatory for and on
behalf of the National Aeronautics Space Administration under contract
NAS8-03060. We also acknowledge the images from the Local Group Survey\footnote{http://www.lowell.edu/$\sim$massey/lgsurvey/}, which were used to
prepare the optical finding charts for the transient sources.


\begin{deluxetable}{ccccc}
\tablecaption{The Properties of the candidate transient X-ray sources}
\tablehead{
\colhead{} &
\colhead{XRT-1} &
\colhead{XRT-2} &
\colhead{XRT-3} &
\colhead{XRT-4} 
}

\startdata
ChASeM33 number & 10 & 130 & 161 & 171\\
R.A. (J2000) & 01 32 41.35 & 01:33:32.23 & 01 33 38.90 & 01 33 41.26\\
Dec. (J2000) & +30 32 18.2 & +30:39:55.8 & +30 21 14.0 & +30 32 13.5\\
Position Error ($''$) & 0.52\tablenotemark{k} & 0.22 & 0.68\tablenotemark{l} & 0.28\\
Total Net Counts\tablenotemark{a}  &   119    &  229 & 34    &   229\\
Photon Index ($\Gamma$) & 0.7$\pm$0.4 & 1.6$\pm$0.4 & \nodata  & 0.6$^{+0.5}_{-0.3}$\\
kT (keV) & \nodata & \nodata & 0.1; & \nodata\\
N$_H$ (10$^{22}$ cm$^{-2}$) &  0.2; & 0.1; & 1; & 0.2$^{+0.6}_{-0.2}$\\
$\chi^2/\nu$ & 1.9 & 1.4 & 0.8 & 0.8\\
DOF & 3 & 10 & 4 & 11\\
Soft\tablenotemark{b} & 2.8$\pm$2.1 & 7.1$\pm$1.3 & 10.6$\pm$5.5 & 3.3$\pm$1.1\\
Med\tablenotemark{c}  & 5.8$\pm$1.1 & 6.2$\pm$0.7 & 8.5$\pm$2.4 & 5.5$\pm$0.7\\
Hard\tablenotemark{d} & 13.8$\pm$2.5 & 6.5$\pm$1.0 & -1.1$\pm$3.1 & 14.4$\pm$1.5\\
Date of Max & 2005-Sep-26 & 2005-Nov-23 & 2005-Oct-01 & 2006-Jun-26\\
Max flux\tablenotemark{e} & 6.6$\pm$1.0 & 2.1$\pm$0.3 & 5.3$\pm$1.2 & 3.5$\pm$0.4\\
Max lum\tablenotemark{f}  & 23 & 6 & 4 & 18\\
Max unabsorbed lum\tablenotemark{g} & 25 & 7 & 830 & 19\\
Min Date & 2006-Jun-15 & 2000-Jul-12 & 2005-Sep-26 & 2005-Nov-21\\
Min flux\tablenotemark{h} & $<$6.6 & $<$2.6 & $<$4.8 & $<$4.1\\
Max/Min\tablenotemark{i} & $>$10 & $>$8 & $>$11 & $>$9\\
Decay Time (days) & $5<t_d<58$ & \nodata & \nodata & \nodata\\
Rise Time (days) & \nodata & \nodata & $t_r<6$ & \nodata\\
m$_V$ & 20.29 &  22.8 & 23.3 & 18.1\\
$B-V$ & 0.03 & 0.11 & -0.1 & -0.2\\
{\sl XMM-Newton}?\tablenotemark{j} & NO & [MPH2006] 145 & [MPH2006] 162? & [MPH2006] 168?\\%
\enddata
\label{table}
\end{deluxetable}

\setcounter{table}{0}

\begin{deluxetable}{cccc}
\tablecaption{The Properties of the candidate transient X-ray sources}
\tablehead{
\colhead{} &
\colhead{XRT-5} &
\colhead{XRT-6} &
\colhead{XRT-7} 
}

\startdata
ChASeM33 Number & 241 & \nodata & 393\\
R.A. (J2000) & 01 34 01.64 & 01 34 09.92 & 01 35 09.11\\
Dec. (J2000) & +30 48 29.8 & +30 32 19.9 & +30 43 41.5\\
Position Error ($''$) & 0.17 & 0.40 & 0.43\\
Counts\tablenotemark{a}  &   226  &   606\tablenotemark{m}  & 53\\
Photon Index ($\Gamma$) & 1.9$^{+0.6}_{-0.4}$ & \nodata & 2.0$^{+2.0}_{-1.3}$\\
kT (keV) & \nodata & 0.054$\pm$0.004 & \nodata\\
N$_H$ (10$^{22}$ cm$^{-2}$) & 0.3$^{+0.3}_{-0.2}$ & 0.17$\pm$0.05 & 0.9$^{+1.5}_{-0.9}$\\
$\chi^2/\nu$ & 1.2 & 2.0 & 0.2\\
DOF & 10 & 22 & 3\\
Soft\tablenotemark{b}  & 7.3$\pm$1.5 & 35.9$\pm$2.7 & 1.0$\pm$1.8\\
Med\tablenotemark{c} & 8.6$\pm$0.9 & 0.0$\pm$0.2 & 5.0$\pm$1.2\\
Hard\tablenotemark{d} & 7.5$\pm$1.3 & 0.4$\pm$0.6 & 5.8$\pm$2.1\\
Date of Max & 2006-Mar-03 & 2000-Aug-02 & 2006-Jun-09\\
Max flux\tablenotemark{e} & 3.1$\pm$0.4 & 158$\pm$13\tablenotemark{n} & 2.1$\pm$0.3\\
Max lum\tablenotemark{f} & 8 & 88 & 3\\
Max unabsorbed lum\tablenotemark{g} & 12 & 620 & 7\\
Date of Min & 2001-Jul-06 & 2006-Jun-26 & 2001-Jul-06\\
Min flux\tablenotemark{h} & $<$2.6 & $<$14.5 & $<$1.9\\
Max/Min\tablenotemark{i} & $>$12 & $>$109 & $>$11\\
Decay Time (days) & \nodata & 1.3 & \nodata\\
Rise Time (days) & \nodata & \nodata & \nodata\\
m$_V$ & 22.1\tablenotemark{o} & \nodata & \nodata\\%
$B-V$ & 0.7\tablenotemark{o} & \nodata & \nodata\\
{\sl XMM-Newton}?\tablenotemark{j} & NO & [MPH2006] 207 (SSS) & NO\\
\enddata 
\end{deluxetable}

\newpage
{\footnotesize
\noindent
$^{\it a}${Net counts summed over all observations of the source.}\\
$^{\it b}${Average 0.35--1.1 keV photon flux from all observations in units of 10$^{-7}$ photons cm$^{-2}$ s$^{-1}$.}\\
$^{\it c}${Average 1.1--2.6 keV photon flux from all observations in units of 10$^{-7}$ photons cm$^{-2}$ s$^{-1}$.}\\
$^{\it d}${Average 2.6--8.0 keV photon flux from all observations in units of 10$^{-7}$ photons cm$^{-2}$ s$^{-1}$.} \\
$^{\it e}${Maximum observed 0.35--8 keV photon flux in units of 10$^{-6}$ photons cm$^{-2}$ s$^{-1}$}\\
$^{\it f}${Maximum observed 0.35--8 keV luminosity, assuming the best-fitting spectral model, in units of 10$^{35}$ erg s$^{-1}$.}\\
$^{\it g}${Maximum observed absorption-corrected 0.35--8 keV luminosity, assuming the best-fitting spectral model, in units of 10$^{35}$ erg s$^{-1}$.}\\
$^{\it h}${Minimum observed 0.35--8 keV photon flux in units of $\times$10$^{-7}$ photons cm$^{-2}$ s$^{-1}$.}\\
$^{\it i}${Lower-limit of the ratio of the maximum 0.35 -- 8 keV flux of the source to the minimum 0.35--8 keV flux of the source.}\\
$^{\it j}${Was the source included in the {\it XMM-Newton} catalogs of \citet{pietsch2004} and/or \citet{misanovic2006}?}\\
$^{\it k}${This random error does not include the effects of the chip edge, which may have skewed the source position as much as 2$''$ to the southeast.}\\
$^{\it l}${This random error does not include the effects of the chip edge, which may have skewed the source position as much as 2$''$ to the northwest.}\\
$^{\it m}${254 counts from Aug. 2 {\it XMM-Newton}; 189 counts from Aug. 4 {\it XMM-Newton}; 163 counts from Aug. 30 {\it Chandra}.}\\
$^{\it n}${Photon flux taken from simultaneous spectral fitting (see \S~3.1).}
\\
$^{\it o}${Magnitudes converted from $g$ and $r$ to $B$ and $V$ using the transformations of \citet{jester2005}.}}\\

\clearpage

\begin{deluxetable}{ccccccc}
\tabletypesize{\footnotesize}
\tablecaption{Magnitudes\tablenotemark{a} and Colors of Mid-Infrared Counterparts to X-ray Transient Candidates in M33}
\tablehead{
\colhead{ID} & \colhead{m$_{3.6}$}  & \colhead{m$_{4.5}$} &  \colhead{m$_{5.8}$}  &  
\colhead{m$_{8.0}$}  &  
\colhead{[3.6]\,$-$[4.5]}   &  
\colhead{[5.8]\,$-$[8.0]}}
\startdata
XRT-1	& \nodata	 &	\nodata	& 	\nodata	& 	\nodata	&	\nodata	&	\nodata\\
XRT-2	& 15.0 $\pm$ 0.2  &	14.8 $\pm$ 0.2 	& 	\nodata	& 	\nodata	&	0.2 $\pm$ 0.3	&	\nodata\\
XRT-3	& \nodata	 &	\nodata	& 	\nodata	& 	\nodata	&	\nodata	&	\nodata\\
XRT-4	& 15.5 $\pm$ 0.3 	 &  15.6 $\pm$ 0.3	& 	\nodata	& 	\nodata	&	$-$0.1 $\pm$ 0.4	&	\nodata\\
XRT-5	&16.2 $\pm$ 0.4	 &	16.0 $\pm$ 0.4 	& 	15.0 $\pm$ 0.5 	& 	\nodata	&	0.2 $\pm$ 0.6	&	\nodata\\
XRT-6	& \nodata	 &	\nodata	& 	\nodata	& 	\nodata	&	\nodata	&	\nodata\\
XRT-7   &  16.9 $\pm$ 0.5  &	15.8 $\pm$ 0.4	&  15.0 $\pm$ 0.3 & 14.0 $\pm$ 0.3 & 1.1 $\pm$ 0.6  & 1.0 $\pm$ 0.5 \\
\enddata
\tablenotetext{a}{Computed according to the Vega-magnitude system as described for IRAC by \citet{reach2005}.}
\label{spitztab}
\end{deluxetable}

\clearpage

\clearpage

\end{document}